\documentclass{aa}
\usepackage{txfonts}
\usepackage{graphics}
\usepackage{latexsym}

\begin{document}
\title{GZ Cancri: a cataclysmic variable \\at the lower edge of the period gap
\thanks{Based on observations made at the European Southern Observatory}}
\author{
C. Tappert\inst{1} \and
A. Bianchini\inst{2}
}
\institute{
Departamento de F\'{\i}sica, Grupo de Astronom\'{\i}a, 
Universidad de Concepci\'on, Casilla 160-C, Concepci\'on, Chile
\and
Dipartimento di Astronomia, Universit\`a di Padova, Vicolo dell'Osservatorio 2,
I-35122 Padova, Italy
}
\offprints{\\C. Tappert, \email{claus@gemini.cfm.udec.cl}}
\date{Received / Accepted}
\titlerunning{
GZ Cnc
}
\authorrunning{Tappert \& Bianchini}

\abstract{
We present calibrated photometry and time-resolved spectroscopy of the 
cataclysmic variable GZ Cnc. Radial velocities of the H$\alpha$ emission line 
reveal an orbital period of 0.08825(28) d, or 2.118(07) h, placing the system 
at the lower edge of the period gap. One of our observations catches the 
system on the rise to an overlooked, possibly small-amplitude, outburst. 
At certain phases during this stage the profiles of the emission lines are
distorted by a high velocity absorption component. We discuss this
phenomenon in the context of the recent suggestion that the long-term
lightcurve of GZ Cnc resembles those of intermediate polars.
}

\maketitle

\section{Introduction}

Cataclysmic variables (CVs) are close interacting binaries with a white dwarf
as primary component and usually a late-type main-sequence star as secondary.
In the course of their evolution from wide detached binaries, angular momentum
loss thought to be driven by magnetic braking and gravitational radiation has
shrunk the Roche lobe of the secondary into contact with the stellar surface.
This causes mass transfer via the Lagrange point $L_1$ into the gravitational
well of the primary. In the absence of strong magnetic fields, an accretion 
disc around the white dwarf is formed, which is also the major source of the
typical emission line spectrum. For a comprehensive overview on CVs see Warner 
(\cite{warn95}).

GZ Cnc (= RX J0915.8+0900 = Tmz V034) was discovered independently as a 
variable star and a ROSAT source (Takamizawa \cite{taka98}; Bade et al.\ 
\cite{bade+98}).  A spectrum was presented by Jiang et al.\ (\cite{jian+00}),
showing a blue continuum and moderately strong emission lines. Although the
spectral coverage went up to 8200 {\AA}, no spectroscopic signatures of
the secondary star were detected. Kato et al.\ (\cite{kato+01}) obtained 
lightcurves of the star during decline in outburst. Their longest time series
spans $\sim$3.1 h and does not show any apparent periodic variations. Both
the spectroscopic and the photometric properties, together with the outburst
behaviour (slow rise, small amplitude; Kato et al.\ \cite{kato+01}) therefore
point to an SS Cyg-type dwarf nova with an orbital period $>$ 3 h. In a recent
paper Kato et al.\ (\cite{kato+02}) investigate the long-term lightcurve in 
more detail, and emphasise similarities to those of intermediate polars. Again,
this would favour a long orbital period, as the vast majority of those
systems is found above the period gap (Ritter \& Kolb \cite{rittkolb98}).

We here present time-resolved spectroscopy in order to derive the orbital 
period, which, in fact, turns out to be less than 3 h. In Sect.\ \ref{obs_sec}
we describe the observations and the reduction process.
The results are presented in Sect.\ \ref{res_sec}  and subsequently 
discussed in Sect.\ \ref{disc_sec}. We complete our paper with a short 
conclusion in Sect.\ \ref{concl_sec}.

\section{Observations and reductions\label{obs_sec}}

\begin{table}
\caption[]{Overview of the observations. $n_{\rm data}$ gives the number of
data points, $t_{\rm exp}$ the individual exposure time, and $\Delta t$ the 
time range covered by the observations.}
\label{obs_tab}
\begin{tabular}{l r r r l}
\hline
\hline
date & $n_{\rm data}$ & $t_{\rm exp}$ [s] & $\Delta t$ [h] & Notes\\
\hline
\multicolumn{4}{l}{\em Spectroscopic data:}\\
2001-01-16 & 12 & 600 & 2.40\\
2001-01-17 & 13 & 600 & 2.48\\
2001-01-18 & 11 & 600 & 2.29\\
2002-03-27 & 16 & 300 & 1.93\\
2002-03-28 & 26 & 300 & 3.01\\
\multicolumn{4}{l}{\em Photometric data:}\\
2001-01-16 & 4 & 20 & 2.26\\
2001-01-17 & 3 & 20 & 2.05\\
2001-01-18 & 4 & 20 & 1.62\\
2002-03-26 & 1 & 60,20,15 & -- & B,V,R\\
2002-03-27 & 7 & 20 & 1.78\\
2002-03-28 & 8 & 20 & 2.36\\ 
\hline
\end{tabular}
\end{table}

The data were taken on January 16--18, 2001, and on March 26--28, 2002, with 
the DFOSC system at the 1.54 m Danish telescope at ESO, La Silla, Chile. The 
spectroscopy was performed using grism \#7, yielding a wavelength range of
3800--7000 {\AA}. A slit width of 1\farcs5 gave a spectral resolution of 4.9 
{\AA}. The measurements were accompanied by He-Ne arc spectra for wavelength 
calibration and by spectrophotometric standards for flux calibration. Between 
each row of spectroscopic measurements, photometric exposures were taken in the
$V$ passband. Additionally, calibrated $BVR$ photometry was obtained
on 2002-03-26. Table \ref{obs_tab} presents the observation log.

The reduction process included bias and flatfield correction, as well as
wavelength and flux calibration, which were performed with IRAF routines
(Tody \cite{tody93}). Further analysis of the spectroscopic data was done with 
ESO-MIDAS (Warmels \cite{warm92}), while the photometry made use of the IRAF 
daophot package and the standalone daomatch and daomaster routines (Stetson
\cite{stet92}). The photometric calibration was obtained by comparison
with Landolt (\cite{land92}).

\section{Results\label{res_sec}}

\subsection{Photometry}

\begin{figure}
\rotatebox{270}{\resizebox{6.0cm}{!}{\includegraphics{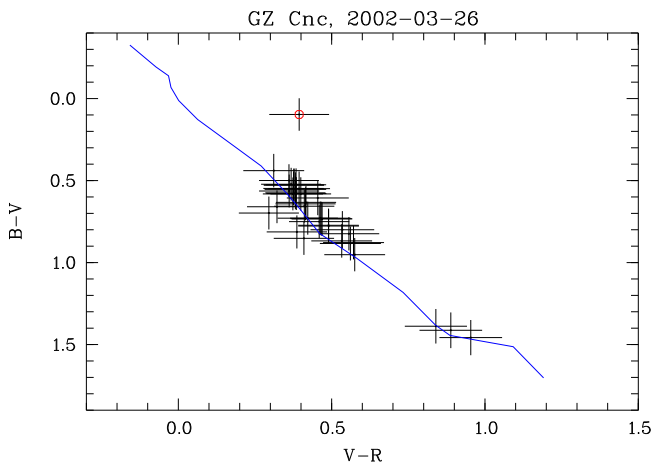}}}
\caption[]{Colour-colour diagram of the field of GZ Cnc, consisting of 38
field stars and GZ Cnc itself (marked by a circle). The range of magnitudes is 
$B = 14.4 - 18.5$, $V = 13.7 - 17.7$, $R = 13.4 - 17.2$. The solid line marks 
the Bessell (\cite{bess90}) main sequence. 
}
\label{col_fig}
\end{figure}

\begin{table}
\caption[]{Calibrated magnitudes for GZ Cnc. Note that the date marks the
start of the night in local time. The errors in Cols.\ 3 and 4 result from the
combination of the photometric error and the error of the differential 
lightcurve, with the exception of the value from 2002-03-26 which represents 
the error of the photometric calibration only.}
\label{ph_tab}
\begin{tabular}{l l l l}
\hline
\hline\noalign{\smallskip}
date & HJD & $V_{\rm min}$ & $V_{\rm max}$ \\
\hline\noalign{\smallskip}
2001-01-16 & 2451926 & 15.460(38) & 15.345(38) \\ 
2001-01-17 & 2451927 & 15.393(38) & 15.228(38) \\
2001-01-18 & 2451928 & 15.144(38) & 14.812(38) \\
2002-03-26 & 2452360 & 15.471(85)\\
2002-03-27 & 2452361 & 15.648(28) & 15.439(27) \\
2002-03-28 & 2452362 & 15.698(27) & 15.357(27) \\
\hline\noalign{\smallskip}
\end{tabular}
\end{table}

\begin{figure}
\rotatebox{270}{\resizebox{5.8cm}{!}{\includegraphics{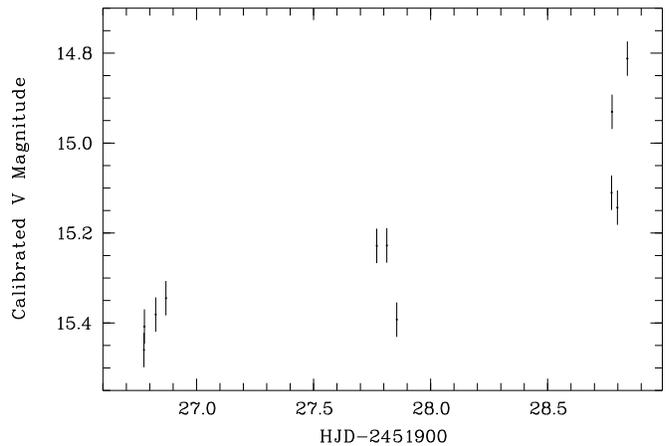}}}
\caption[]{Photometric data of GZ Cnc for the 2001 run.}
\label{mag01_fig}
\end{figure}

Only on 2002-03-26 direct calibrated photometric data were obtained in $B$,
$V$, and $R$ passbands. The resulting colours are
\begin{eqnarray*}
B-V = 0.097(99), & ~V-R = 0.394(97), \\
{\rm at}~V = 15.471(85), 
\end{eqnarray*} 
which lie in the usual range occupied by dwarf novae (Echevarr\'{\i}a
\cite{eche84}), and put it clearly away from the main sequence 
(Fig.\ \ref{col_fig}).

For the other nights, calibrated $V$ magnitudes for GZ Cnc were computed
with respect to an average lightcurve composed of 5 stars which could be
measured in all frames and proved to be constant within the errors. The 
resulting magnitudes are given in Table \ref{ph_tab}. During the 2002 observing
run the system appears to have been in a stable quiescent state, with
variations $\Delta m_V \sim 0.3$ mag probably due to flickering or an 
additional continuum source (e.g., the bright spot). A comparison with Kato et 
al.\ (\cite{kato+02}) shows that our observations lie close to two recorded 
outbursts on 2002-03-11 and 2002-04-02. The 2001 data, however, clearly show a 
brightness increase, which could be interpreted as the onset of an 
outburst (see Fig.\ \ref{mag01_fig}), although to our knowledge none has been 
registered in the subsequent days. However, the coverage of GZ Cnc contains a 
lot of gaps. For the year 2001, only 11 observations are recorded in the VSNET 
database\footnote{The archive of the VSNET mailing list is located at 
http://www.kusastro.kyoto-u.ac.jp/vsnet/Mail}, among them one single positive 
detection. A weak outburst could thus easily have been missed.

The increase rate of roughly 0.2 mag d$^-1$ in our data fits well with the
outburst recorded by Kato et al.\ (\cite{kato+01}), that showed a rate of
$\sim$0.25 mag d$^-1$ over 5 nights. This lends further support to the idea
of an overlooked outburst in 2001.

\subsection{Properties of the average spectra}

\begin{figure}
\resizebox{8.0cm}{!}{\includegraphics{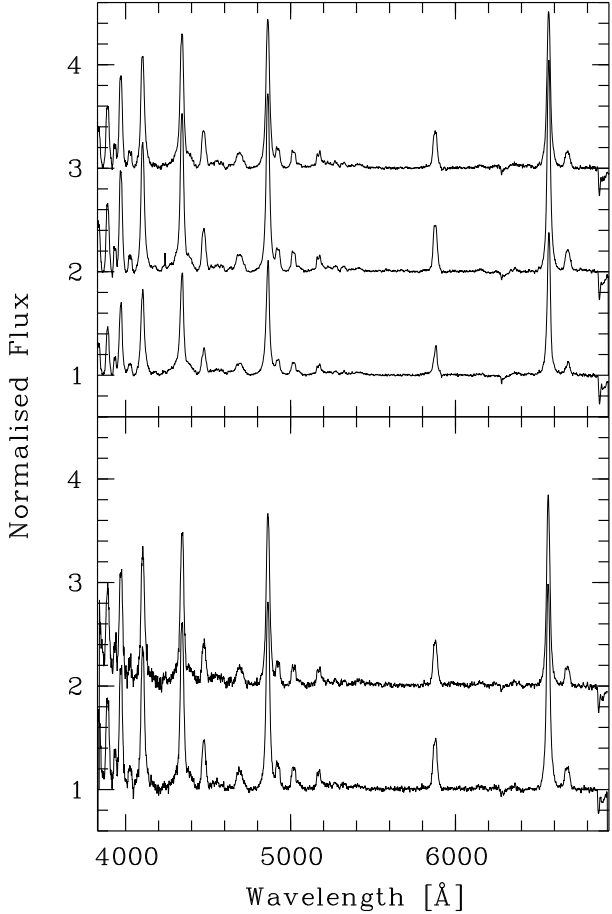}}
\caption[]{Average normalised spectra of GZ Cnc. {\bf Above:} Spectra from
2001-01-16, 2001-01-17, and 2001-01-18 (from top to bottom). The former two
have been displaced by 2 and 1 units in vertical direction. The small spike at 
$\lambda =$ 4240 {\AA} in the 2001-01-17 spectrum is an artifact, the 
absorption lines in the red part of the spectrum are atmospheric. {\bf Below:}
Spectra from 2002-03-27 (top, displaced vertically by 1 unit) and 2002-03-28 
(bottom).}
\label{navsp_fig}
\end{figure}

\begin{table}
\caption[]{Identified emission lines and their equivalent widths. Only lines
which were detected in all five averaged spectra are shown. Col.\ 1 gives the 
average wavelength position of the lines in the spectra, Col.\ 2 the 
identification, and Col.\ 3 the corresponding rest wavelength. Cols.\ 4-8 
contain the measured equivalent widths for the five nights. In two cases the 
values for the blend of two lines are given for the stronger line: 
\ion{Ca}{II}$\lambda$3934+H$\epsilon$ (note that H$\epsilon$ itself already 
includes the other \ion{Ca}{II} line) and H$\beta$+\ion{He}{I}$\lambda$4922.}
\label{eqw_tab}
\begin{tabular}{l l l r r r r r}
\hline
\hline\noalign{\smallskip}
$\lambda$ [{\AA}] & ID & $\lambda_0$ [{\AA}] & 
\multicolumn{5}{c}{$W_{\lambda}$ [{\AA}]} \\
 & & & \multicolumn{3}{c}{~2001-01} & \multicolumn{2}{c}{~~2002-03} \\
 & & & 16 & 17 & 18 & 27 & 28 \\
\hline
3890.5 & H$\zeta$      & 3889.1 & $-$15 & $-$17 & $-$10 & $-$15 & $-$16 \\ 
3935.9 & \ion{Ca}{II}  & 3933.7 &       &       &       &       &       \\ 
3971.0 & H$\epsilon$   & 3970.1 & $-$26 & $-$30 & $-$18 & $-$28 & $-$28 \\
4026.0 & \ion{He}{I}   & 4026.2 &  $-$3 &  $-$4 &  $-$2 &  $-$3 &  $-$5 \\
4103.4 & H$\delta$     & 4101.7 & $-$33 & $-$36 & $-$24 & $-$35 & $-$44 \\
4341.7 & H$\gamma$     & 4340.5 & $-$45 & $-$46 & $-$38 & $-$31 & $-$53 \\
4474.1 & \ion{He}{I}   & 4471.5 & $-$11 & $-$12 &  $-$7 &  $-$9 & $-$13 \\
4688.9 & \ion{He}{II}  & 4685.7 & $-$10 &  $-$8 &  $-$7 &  $-$9 &  $-$8 \\
4862.4 & H$\beta$      & 4861.3 & $-$53 & $-$60 & $-$39 & $-$59 & $-$64 \\
4920.4 & \ion{He}{I}   & 4921.9 &       &       &       &       &       \\
5018.1 & \ion{He}{I}   & 5015.7 &  $-$6 &  $-$7 &  $-$5  & $-$8 &  $-$8 \\
5172.2 & \ion{Fe}{II}  & 5169.0 &  $-$4 &  $-$4 &  $-$3 &  $-$4 &  $-$5 \\
5268.7 & \ion{Fe}{II}  & 5272.4 &  $-$1 &  $-$1 &  $-$1 &  $-$1 &  $-$1 \\
5321.4 & \ion{Fe}{II}  & 5325.6 &  $-$1 &  $-$1 &  $-$1 &  $-$2 &  $-$1 \\
5876.7 & \ion{He}{I}   & 5875.6 & $-$12 & $-$14 &  $-$8 & $-$15 & $-$16 \\
6563.9 & H$\alpha$     & 6562.8 & $-$46 & $-$59 & $-$38 & $-$53 & $-$58 \\
6677.8 & \ion{He}{I}   & 6678.1 &  $-$5 &  $-$7 &  $-$3 &  $-$6 &  $-$6 \\
\hline
\end{tabular}
\end{table}

The spectrum of GZ Cnc (Fig.\ \ref{navsp_fig}) resembles that of a dwarf nova, 
with dominating emission lines from the Balmer and \ion{He}{I} series. Further
lines include \ion{He}{II} $\lambda$4686, as well as \ion{Fe}{II} 
$\lambda$5172. Table \ref{eqw_tab} summarises the identified lines
and their equivalent widths. The emission lines are of rather moderate 
strength (e.g., Williams \cite{will83}), which indicates a moderately high
mass-transfer rate (Fig.\ 6 in Patterson \cite{patt84}). As noted by
Kato et al.\ (\cite{kato+02}), \ion{He}{II} appears to be somewhat stronger 
than in most dwarf novae, and especially those with a short orbital period.

The slow rise in the photometry during the 2001 run is also roughly reflected 
by the equivalent widths of the emission line, which first show a slight rise 
from the first to the second night, followed by a steep decline on the third 
night (Table \ref{eqw_tab}, Fig.\ \ref{navsp_fig}). It appears thus that this 
variation is indeed due to a brightening of the disc continuum. 

\subsection{Period analysis}

\begin{figure}
\resizebox{8.0cm}{!}{\includegraphics{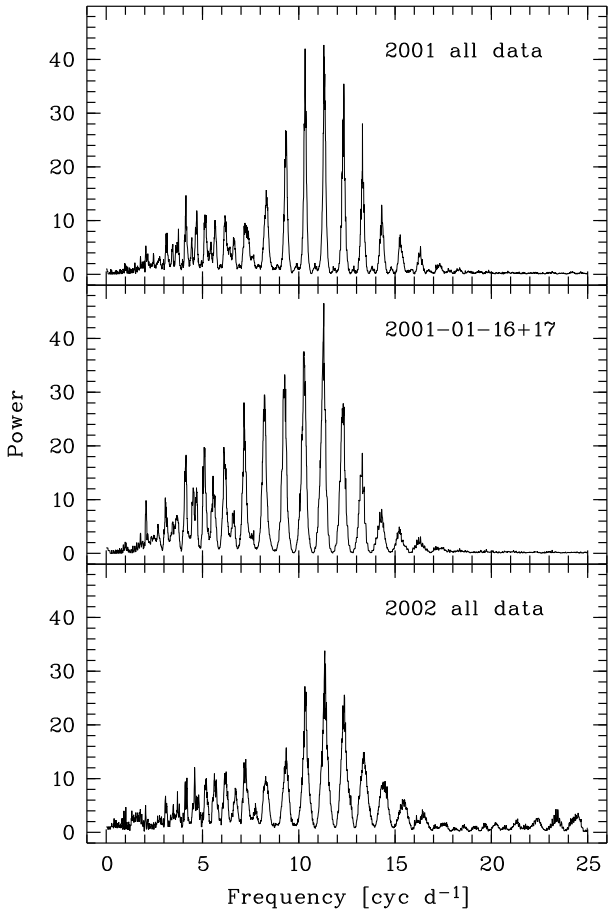}}
\caption[]{AOV periodograms for the data sets as indicated in the plot. The
highest peak in all periodograms corresponds to $P = 2.12$ h.}
\label{pergram_fig}
\end{figure}

\begin{table*}
\caption[]{The three principal frequencies and their 
statistical quantities $p_c$ and $p_d$ as explained in the text.}
\label{freq_tab}
\begin{tabular}{l l l l l l l l l l}
\hline
\hline\noalign{\smallskip}
 &
\multicolumn{3}{c}{2001 all data} &
\multicolumn{3}{c}{2001-01-16+17} &
\multicolumn{3}{c}{2002 all data} \\
 & $f$ [cyc d$^{-1}$] & $p_c$ & $p_d$ &
   $f$ [cyc d$^{-1}$] & $p_c$ & $p_d$ &
   $f$ [cyc d$^{-1}$] & $p_c$ & $p_d$ \\
\hline\noalign{\smallskip}
$f_1$ & 10.343(39) & 0.344 & 0.381 &  
        10.293(46) & 0.383 & 0.360 & 
        10.346(42) & 0.252 & 0.246 \\
$f_2$ & 11.340(41) & 0.463 & 0.432 &  
        11.301(40) & 0.532 & 0.538 & 
        11.362(61) & 0.653 & 0.597 \\
$f_3$ & 12.336(37) & 0.192 & 0.187 & 
        12.293(53) & 0.084 & 0.102 & 
        12.344(48) & 0.095 & 0.157 \\
\hline\noalign{\smallskip}
\end{tabular}
\end{table*}

\begin{figure}
\rotatebox{270}{\resizebox{5.8cm}{!}{\includegraphics{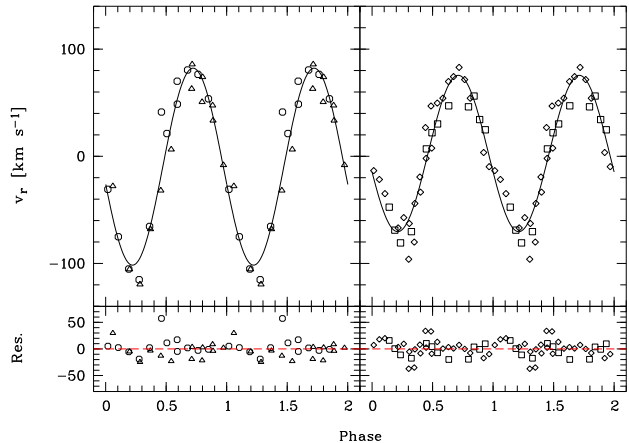}}}
\caption[]{Radial velocities folded on the best period $P$ = 0.08825 d. The 
left plot shows the 2001-01-16 ({\large $\circ$}) and 2001-01-17 data 
($\bigtriangleup$), the 2002-03-27 ($\Box$) and 2002-03-28 
({\large $\diamond$}) velocities are presented in the right plot. The 
respective phase zero points correspond to the values derived with the
diagnostic diagram (Sect.\ \ref{lineprof_sec}). The best sine fit and the fit 
residuals are also shown.}
\label{rvs_fig}
\end{figure}

The Doppler variation of the emission lines caused by the orbital motion
of the system has been measured with a Gaussian fit to the whole H$\alpha$ line
profile. Radial velocities were computed with respect to the \ion{O}{I} night 
sky emission line at $\lambda$5577 {\AA}. As the mean velocity showed strong
nightly variations, and the respective time ranges proved to cover more than 
one orbit, we normalised the velocity data individually by subtracting the
respective mean velocities.

The search for periodicities was conducted using the Scargle (\cite{scar82})
and AOV (Schwarzenberg-Czerny \cite{schw89}) algorithms as implemented in 
MIDAS' tsa context. Unfortunately, the time span between both observing runs
proved to be too long for an analysis of the combined data, so that the runs 
had to be investigated individually. The Scargle algorithm produced five
significant peaks in the range $f = 9.35 - 13.35~\mathrm{cyc~d}^{-1}$. The 
frequencies corresponding to the highest peaks were 10.34 for the 2001 data and
11.34 for 2002. The AOV method in both cases picked out $f \approx 11.35$. For 
the 2002 data, this peak is well separated from the next highest one, for 2001 
the two peaks of 10.34 and 11.34 are very close (Fig.\ \ref{pergram_fig}, lower
and upper plot). 

In most CVs, additional emission components distort the symmetric, disc-borne,
line profile (Tappert \& Hanuschik \cite{tapphanu01}), with the type and 
strength of this distortion being likely to depend on the luminosity state
of the system (e.g., Tappert \cite{tapp99}). Since on 2001-01-18 both the
photometric magnitudes (Table \ref{ph_tab}) and the emission line equivalent
widths (Table \ref{eqw_tab}) differ strongly from those of the other nights in
this run, we also conducted a period analysis for the data of the first two 
nights in 2001 only. Indeed the peak at $f = 11.3$ in the AOV periodogram is 
now much more pronounced with respect to the other frequencies 
(Fig.\ \ref{pergram_fig}, middle plot). 

In order to test this period against the two most prominent aliases at 10.3 and
12.3 cyc d$^-1$ (referring to the 2002-03 data) we have run a Monte Carlo 
simulation (Mennickent \& Tappert \cite{menntapp01}). This routine computes a 
thousand data sets by adding a random value to each data point, based on the 
variance of the sine fit to the real data. A periodogram for each of these sets
is established and the frequencies and their power are recorded. Equivalently 
to the method outlined by Thorstensen \& Freed (\cite{thorfree85}; hereafter 
TF85) one can now compute a {\em discriminatory power} $p_d$ by dividing the 
number of times where a specific frequency has the largest power in comparison 
to the other aliases by the number of total tries. The second statistical 
measure used by TF85, the {\em correctness likelihood} $p_c$, can be obtained
by counting the number of times when the ratio of the power of the highest peak
and that of the second highest one for the artificial data corresponds to that 
of the real data within a certain range of ratios (TF85 propose a typical value
of 0.1). 

Table \ref{freq_tab} summarises the results of this analysis for the three
most prominent frequencies. The highest statistical significance for the
principal frequency $f_2 \sim 11.3$ cyc d$^{-1}$ is obtained to $p_c$ = 0.65 in
the 2002-03 data. This favours the identification of $f_2$ with the orbital 
frequency, but is certainly too low to provide an unambiguous decision, with
$f_1 \sim 10.3$ cyc d$^{-1}$ representing an important alias. 

After the initial submission of this paper we have been informed of the
research by Fenton \& Thorstensen (private communication) who obtained
$P = 0.0880(2)$ d from time-resolved spectroscopy of GZ Cnc. They report that 
their run spans over 6 h of hour angle and does not present any daily alias 
problems.

This period fits very well with our best frequency $f_2$, which, as an average
of the `quiescence' data sets, corresponds to a period
\begin{displaymath}
P = 0.08825(28)~\mathrm{d}~= 2.118(07)~\mathrm{h}~.
\end{displaymath}
We therefore identify this period with the orbital variation.

Fig.\ \ref{rvs_fig} presents the radial velocities folded on our period.

\subsection{Line-profile analysis\label{lineprof_sec}}

\begin{figure}
\resizebox{8.0cm}{!}{\includegraphics{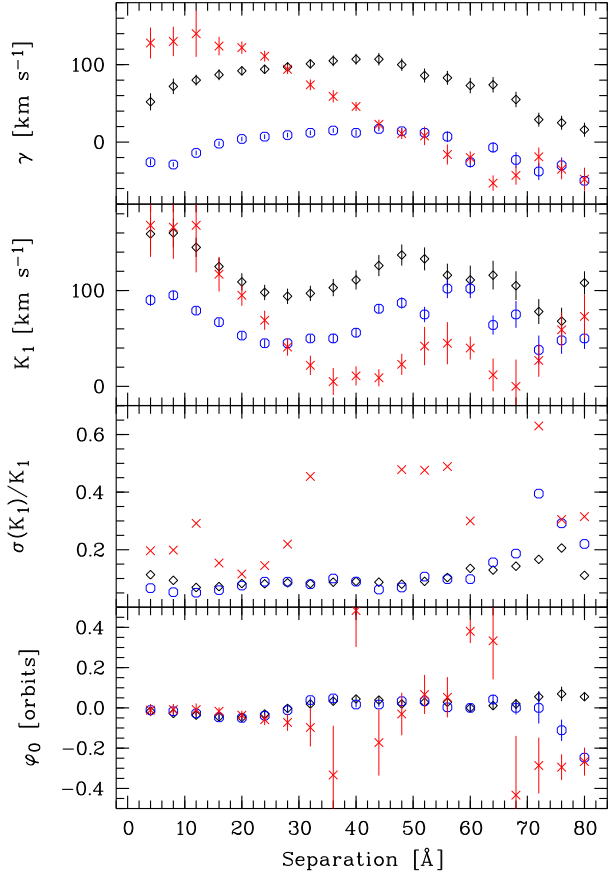}}
\caption[]{Diagnostic diagrams for the following data sets: 2001-01-16+17
($\Diamond$), 2001-01-18 ($\times$), and 2002-03 ({\Large $\circ$}). The zero 
point of $\varphi_0$ was calculated with respect to the value corresponding to
a separation $d$ = 60 {\AA} for each observing run individually (i.e.,
2001-01-16+17 and 2001-01-18  were corrected for the 2001-01-16+17 value,
the 2002-03 set for the 2002-03 value). }
\label{dd_fig}
\end{figure}

\begin{figure}
\resizebox{8.0cm}{!}{\includegraphics{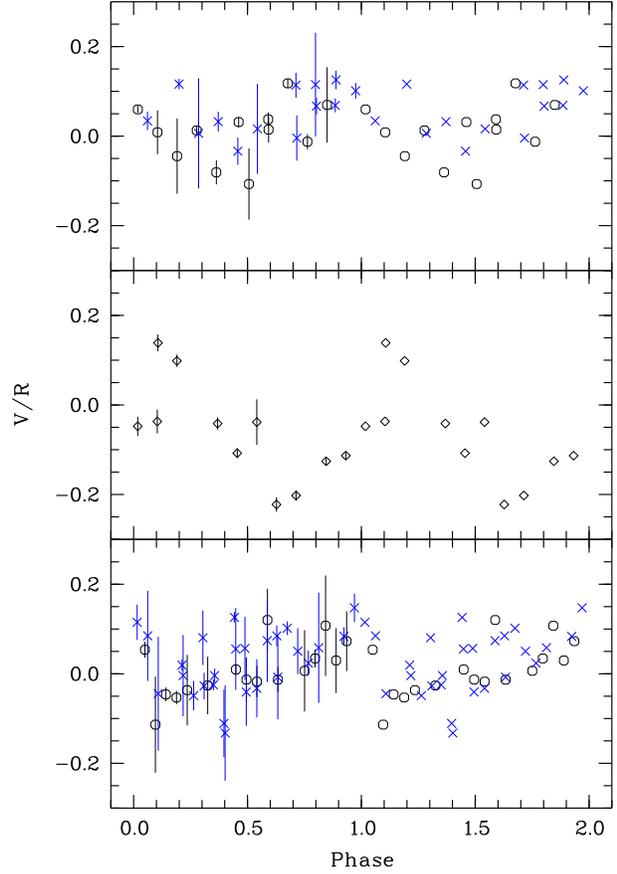}}
\caption[]{$V/R$ plots for the individual data sets. Top: 2001-01-16 
({\Large $\circ$}) and 2001-01-17 ($\times$). Middle: 2001-01-18. Bottom:
2002-03-27 ({\Large $\circ$}) and 2002-03-28 ($\times$). Two phase cycles have
been plotted, the first one including errors. The latter have been determined
by a Monte Carlo simulation. Note that the 2001 data share a common zero
phase.}
\label{vr_fig}
\end{figure}

\begin{figure}
\resizebox{8.0cm}{!}{\includegraphics{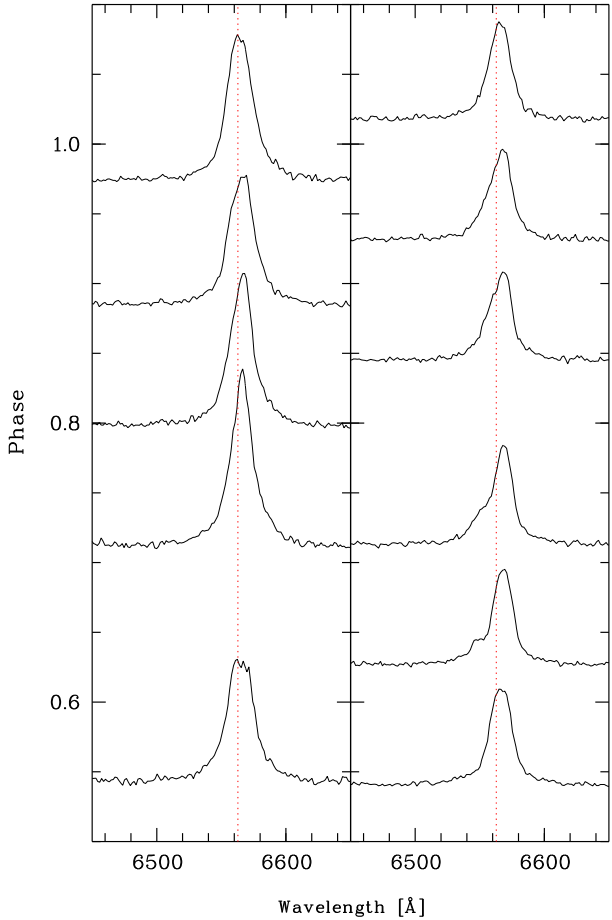}}
\caption[]{Selected line profiles between phases 0.5 and 1.0 for the
2001-01-17 (left) and the 2001-01-18 data (right). The spectra have been
continuum normalised, divided by 20, and shifted vertically according to
phase. The dotted line marks the rest wavelength of H$\alpha$.}  
\label{profile_fig}
\end{figure}

Even for medium spectral resolution an analysis of the emission line profile is
always worth undertaking, as at least it might provide a motivation for
further studies. In the present case, the fact that at least one data set 
caught the system in a different luminosity state, and that its exclusion led 
to a more unambiguous determination of the orbital period, already calls for a 
more thorough investigation. The only really feasible line in our data for such
a study is H$\alpha$, as other lines are either too noisy (\ion{He}{I}, 
\ion{He}{II}, etc.) or are distorted by blends (basically all Balmer lines from
H$\beta$ on bluewards).

Due to the modest spectral and phase resolution the more advanced technique of
Doppler tomography (Marsh \& Horne \cite{marshorn88}), which in principle
would yield the clearest picture on the emission distribution in the system,
does not provide any significant results. However, the application of the
diagnostic diagram (Shafter \cite{shaf83}) and the $V/R$ plot (Tappert
\cite{tapp99}) will still give at least a rough idea of a possible additional
component and of the differences between the individual data sets. 
 
In the former method, two identical Gaussians (in the present case with
FWHM = 4 {\AA}), horizontally separated by a quantity $d$, are used to measure 
radial velocities of different parts of the emission line by variation of $d$
(Schneider \& Young \cite{schnyoun80}). The resulting velocities 
$v(d, \varphi)$ are fit with a function
\begin{displaymath}
v(d, \varphi) = \gamma(d) - K_1(d)~\sin(\varphi).
\end{displaymath}
The fit parameters are plotted as a function of $d$ to yield the
diagnostic diagram. In order to identify the part of the line which is
undisturbed from possible additional components, one usually also plots 
the percentage error of the semi-amplitude $K_1$, $\sigma(K_1)/K_1$. 
A sharp increase in the latter at higher separations can be interpreted as
the continuum noise dominating over the signal from the emission line. The 
last separation before this happens, $d_{\rm max}$ is usually taken as 
that part of the line that reflects best the motion of the white dwarf, 
especially if the corresponding parameters tend to approach constant values. 
Note, that this assumes that the wings of the line profile stem from material 
which is symmetrically distributed within the accretion disc. 

With the comparatively low resolution of our data we do not expect to be able
to extract significant parameters from the diagnostic diagram, but are mainly
interested in comparing the behaviour of the individual data sets. We have 
first computed diagrams for each night separately. In the case of the sets
from 2001-01-16 and 2001-01-17, and in the case of 2002-03-27 and 2002-03-28,
the results were identical within the errors, and we therefore used the
combined data sets. The resulting diagrams are shown in Fig.\ \ref{dd_fig}.
The $\sigma(K_1)/K_1$ curves do not really give a good indication for 
$d_{\rm max}$, and thus our choice to correct the zero point of the phase
for the value corresponding to $d$ = 60 {\AA} is somewhat arbitrary. We were
originally motivated by the behaviour of the 2002-03 $K_1$ and 
$\sigma(K_1)/K_1$ curves, and for the 2001-01-16+17 data by the fact that
at this point the $K_1$ value corresponds well to the 2002-03 $K_1$. However,
we are not particularly interested in a specific value for $\gamma$ or $K_1$,
and although $\varphi_0$ is important for the interpretation of the 
subsequent $V/R$ plot, it stays almost constant over the whole range of
separations, and its choice therefore does not appear to be critical. 

From the appearance of the curves in the diagnostic diagram we can derive
two pieces of information. First, that the variation of $K_1$ provides 
evidence for the presence of an additional emission component, and second,
that the similarity of the variation in all parameters of the 2001-01-16+17 
and 2002-03 data sets indicates a similar type of additional emission, while
there are strong differences with respect to the 2001-01-18 data. 

For the $V/R$ plot the fluxes of the blue (violet) half, $F(V)$, and of the
red half, $F(R)$, of the emission line profile are computed to obtain
\begin{displaymath}
V/R = \log \frac{F(V)}{F(R)}.
\end{displaymath}
The outer borders of the line profile are defined by a certain intensity level
$I$, the central wavelength which separates both halves is taken as the
mean of the two outer wavelengths $\lambda_V(I)$ and $\lambda_R(I)$. For the
present case we chose $I = 0.15~I_{\rm max}$, with $I_{\rm max}$ being the
peak value of the emission line.

The resulting plots, $V/R$ vs.\ phase, are given in Fig.\ \ref{vr_fig}. While
the variation is very noisy and badly defined in the 2001-01-16 and 17, and the
2002-03 data, they do show a similar behaviour. In both cases a 
maximum at similar phases (0.9 in the upper plot, 1.0 in the lower one)
can be identified. As the respective zero points of the phases were determined
individually, we do not consider the observed phase difference as evidence
for a fundamentally different line profile variation. The plot for the
2001-01-18 data (middle plot), however, shows a much stronger, better defined,
variation, which additionally is significantly shifted in phase. 

Maximum and minimum in the $V/R$ plot can usually be identified as the maximum
blue- and redshift of an additional emission component. This assumes that this
component is dominating the line profile and mostly confined to the centre
of the profile. However, this does not seem to work for GZ Cnc. Fig.\ 
\ref{profile_fig} shows that the minimum in the 2001-01-18 data around phase 
0.65 is due to a small bump in the blue wing which then moves redwards with 
phase. Note that the stronger part of the line profile
shows very similar velocities in both the 2001-01-18 and the 2001-01-17 data.

\begin{figure*}
\rotatebox{270}{\resizebox{11.6cm}{!}{\includegraphics{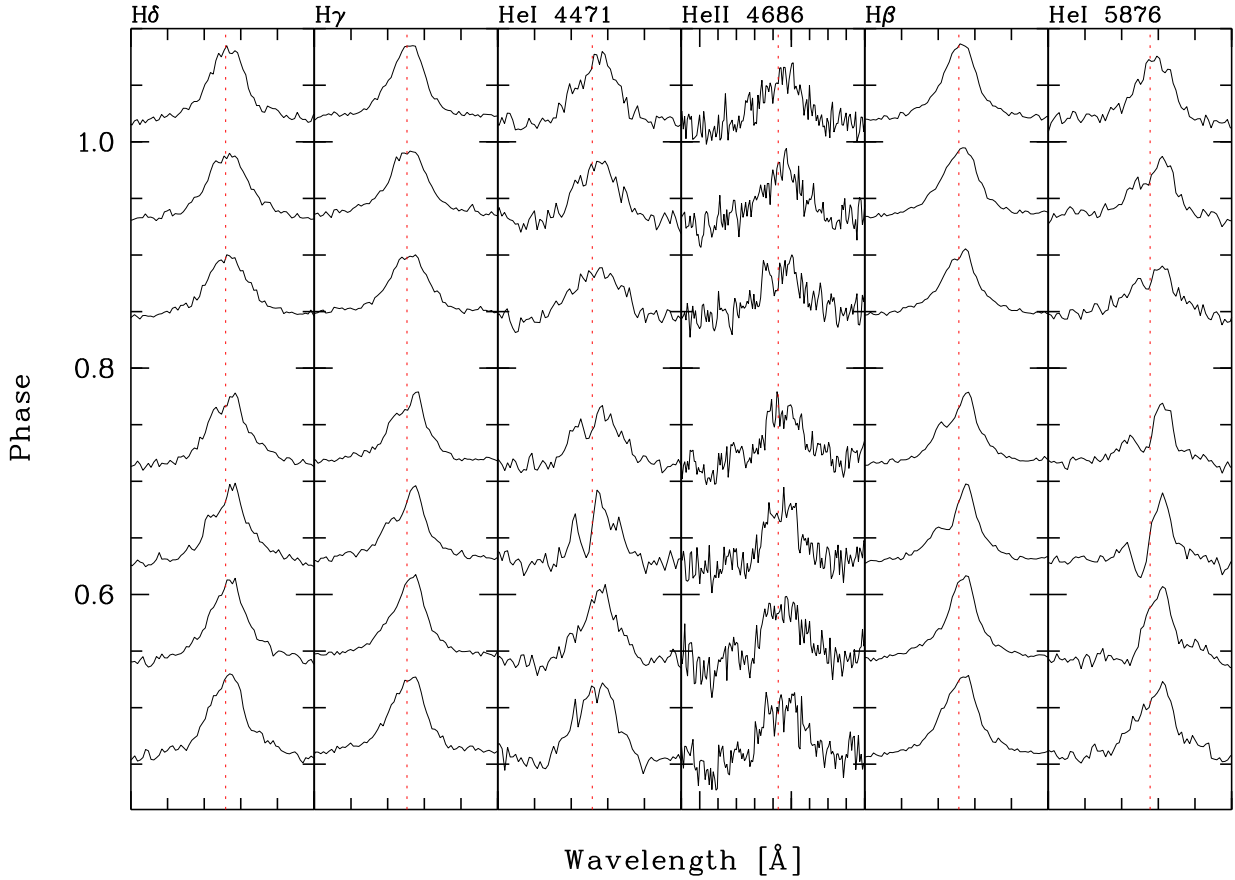}}}
\caption[]{Several line profiles, as indicated in the plot, for selected phases
of the 2001-01-18 data. Like in Fig.\ \ref{profile_fig}, the spectra have 
been continuum normalised, divided by a certain quantity, and shifted 
vertically according to phase. The dotted lines mark the respective rest 
wavelengths. Each unit in the wavelength axis marks 20 {\AA}. Note that
the wavelength range for each plot amounts to 100 {\AA}, except for the much 
broader \ion{He}{II} line, where 200 {\AA} are covered.}  
\label{lines_fig}
\end{figure*}

\begin{figure}
\rotatebox{270}{\resizebox{5.8cm}{!}{\includegraphics{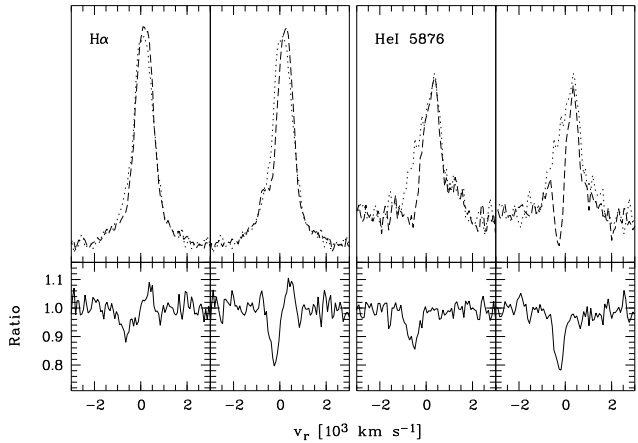}}}
\caption[]{H$\alpha$ and \ion{He}{I} $\lambda$5876 line profiles of phases
0.54 (dashed, left plots of the respective line) and 0.63 (dashed, right plots)
and 0.45 (dotted) on 2001-01-18. The lower plots give the ratio of the later 
phases with respect to phase 0.45.}
\label{HaHecomp_fig}
\end{figure}

While this additional component in H$\alpha$ at first glance appears to be an 
emission feature, a closer investigation of all lines (Fig.\ \ref{lines_fig})
and a comparison with respect to unaffected phases (Fig.\ \ref{HaHecomp_fig})
reveals that it is actually a high velocity absorption component. It is visible
in all emission lines from phase 0.54 to 0.93, with the probable exception of 
\ion{He}{II}.

\begin{figure}
\rotatebox{270}{\resizebox{5.8cm}{!}{\includegraphics{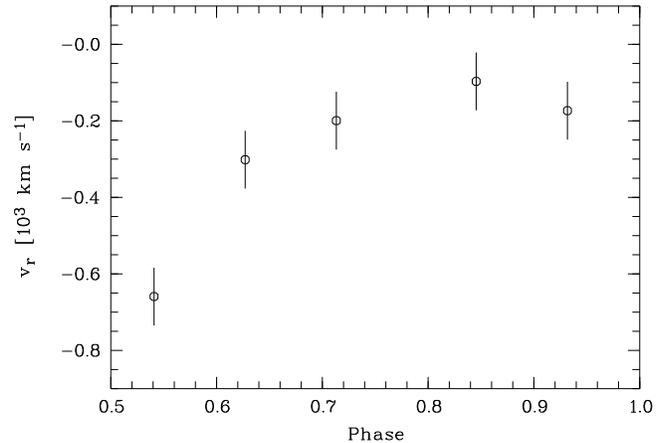}}}
\caption[]{Radial velocities of the \ion{He}{I} $\lambda$5876 absorption 
component as measured with a graphics cursor. The error bars refer to $\pm$1 
unit of the dispersion (76 km s$^-1$), with the real uncertainty being
probably higher at most phases.}
\label{iesvel_fig}
\end{figure}

\begin{figure}
\rotatebox{270}{\resizebox{5.8cm}{!}{\includegraphics{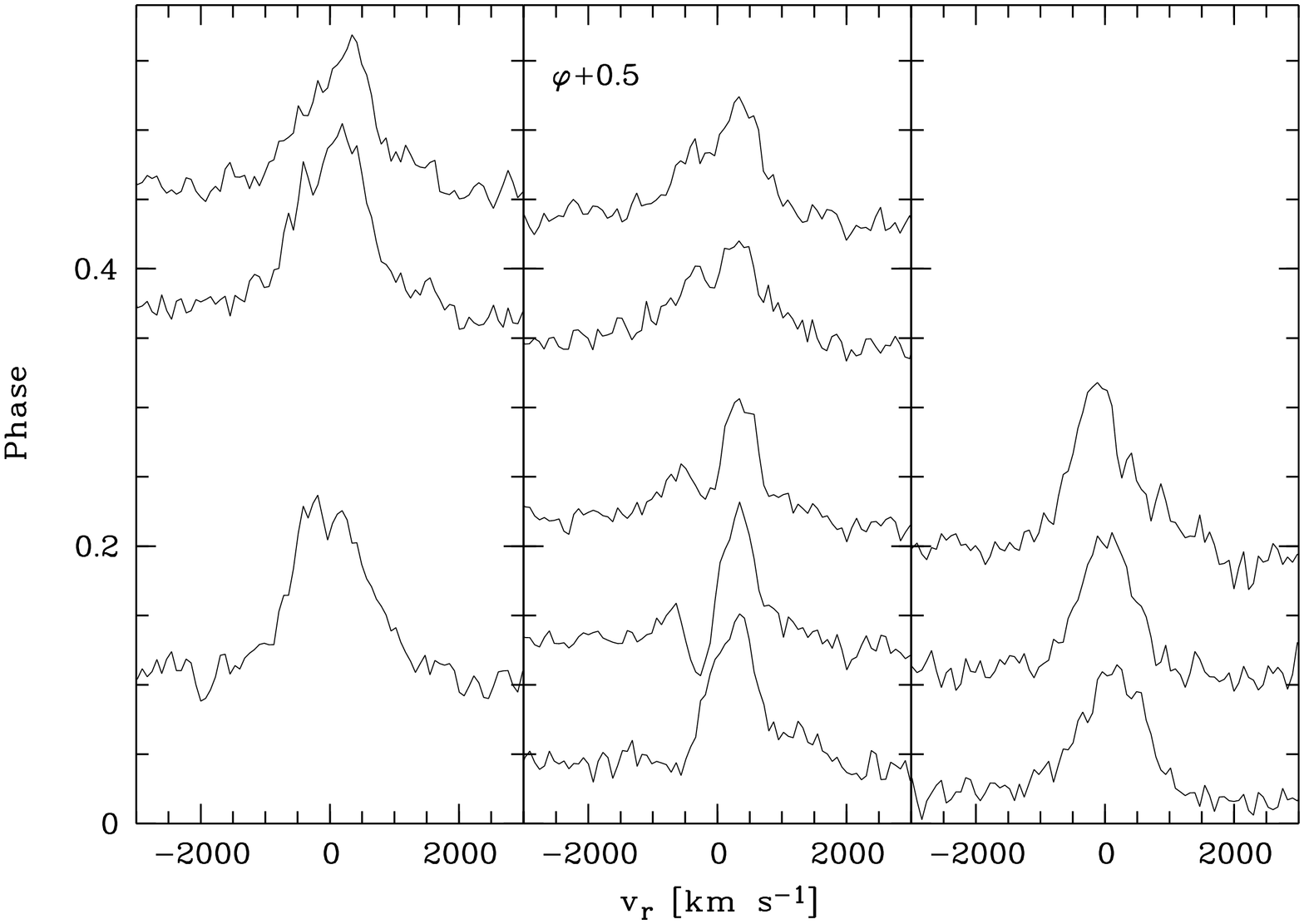}}}
\caption[]{All 2001-01-18 \ion{He}{I} $\lambda$5876 line profiles in time
order from lower left to upper right. Phases between 0.5 and 1.0 are shown
in the middle plot in order to facilitate comparison with opposite phases.}
\label{HeI_fig}
\end{figure}

As the component is most clearly seen in the \ion{He}{I} $\lambda$5876 line, we
chose the latter to measure its velocities. Fig.\ \ref{iesvel_fig} shows that 
it appears at phase 0.54 at a velocity $\sim -650$ km s$^{-1}$, moving redwards
and apparently reaching a plateau around phase 0.9. The latter would suggest
a confinement to negative velocities, but this impression is mainly caused by
the first and the last data point, which certainly are the most uncertain ones
(see Fig.\ \ref{lines_fig}). On the other hand, we find no evidence that the
component ever appears at positive velocities (Fig.\ \ref{HeI_fig}).
Unfortunately, our data does not cover a second orbit, so that it remains
unclear if the absorption component represents a transient or a periodic
event.

\section{Discussion\label{disc_sec}}

\begin{figure}
\rotatebox{270}{\resizebox{5.8cm}{!}{\includegraphics{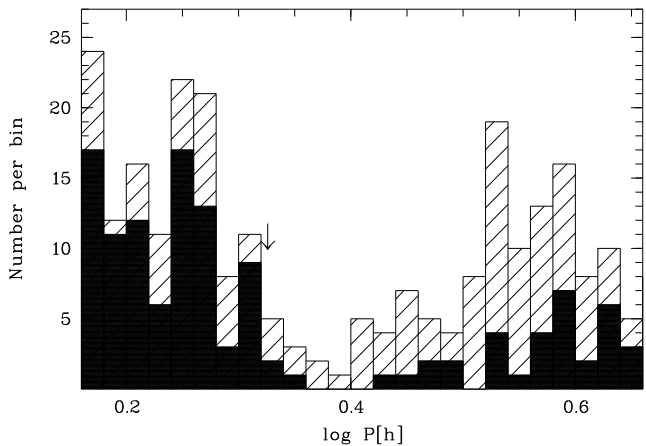}}}
\caption[]{The period distribution of CVs around the period gap. The hashed
histogram represents the distribution of all CVs, the solid one those of
dwarf novae only. The data were taken from the TPP catalogue (Kube et al.\
\cite{kube+02}) as of 2002-11-05. The arrow marks the position of GZ Cnc, which
is not included in either histogram.
}
\label{perdis_fig}
\end{figure}

Our analysis revealed a period of 2.118(7) h, placing GZ Cnc just at the lower 
edge of the period gap of CVs (Fig.\ \ref{perdis_fig}). This is a region which 
is sparsely populated, especially by dwarf novae, and it is therefore of 
particular interest to determine the CV subtype of the system. 

Kato et al.\ (\cite{kato+02}; hereafter K2002) studied the long-term behaviour,
and found a similarity to those of intermediate polars (IPs) in that the 
outburst distribution appeared to be `clustered'. Specifically the authors 
mention the detection of four outbursts within two months in 2002, while the 
only known outburst before that date has been recorded at the end of 2000. 
Still, our data proves that at least one brightening has occurred during 2001 
that remained undetected. It is therefore possible that the observed clustering
is not due to a variation of the outburst frequency, but caused by a variable 
outburst amplitude, especially as the upper detection limit in the observations
used by K2002 varies between $m_{\rm vis} \approx$14.8 and 13.2, the latter 
being already very close to the object's maximum intensity (Fig.\ 1 in K2002). 
Note, however, that frequent outbursts of variable amplitude by no means 
contradict the suggested similarity to \object{V426 Oph}: it certainly fits 
well with the latter's long-term lightcurve (Fig.\ 2 in K2002).

Another point in favour of the IP scenario might be provided by the appearance 
of the absorption component during rise to outburst. Such features during
certain phases have been observed e.g.\ in early outburst of the short period 
($P_{\rm orb}$ = 1.63 h) IP \object{EX Hya} (Hellier et al.\ \cite{hell+89})
and in the long period ($P_{\rm orb}$ = 4.85 h) IP \object{FO Aqr} (Hellier et 
al.\ \cite{hell+90}). They are furthermore a common property of the SW Sex 
stars (Hellier \cite{hell00}, and references therein), which have been recently
proposed to represent high mass-transfer IPs by Rodr\'{\i}guez-Gil et al.\ 
(\cite{rodr+01}). 

These features have been explained by a scenario where part of the gas stream 
from the secondary overflows the disc, and impacts on the inner part of the
disc or connects with the magnetosphere of the white dwarf (Hellier 
\cite{hell00}; Hellier et al.\ \cite{hell+89}). However, an important 
difference of the GZ Cnc absorption with respect to the above systems is that 
their absorption features consist of low-velocity components. A stellar wind
as suggested for a similar phenomenon in the nova-like \object{BZ Cam} 
(Patterson et al.\ \cite{patt+96}) might represent a more probable alternative.

Another similarity with EX Hya, but also a common feature of dwarf novae, lies 
in the presence of an isolated emission component in the quiescence data, 
as suggested by the $V/R$ plot (Fig.\ \ref{vr_fig}). This component is confined
to the line centre and could thus stem from the impact region of the gas stream
with the outer accretion disc.

\section{Conclusions\label{concl_sec}}

There are three pieces of evidence which might favour GZ Cnc as an intermediate
polar (IP): the long-term behaviour (Kato et al.\ \cite{kato+02}), the
unusual, although not exorbitant, strength of the \ion{He}{II} line, and
the appearance of the absorption component during rise to outburst.
None of these is really conclusive, basically because they are based on
insufficient data. However, in their sum they represent at least a strong 
indication.

We conclude that further observations are necessary in order to decide pro or
contra an IP nature for GZ Cnc. Especially helpful would be a better sampled
long-term lightcurve and high-speed photometry. If an IP nature were to be
confirmed, it would make GZ Cnc a rare system indeed: the TPP catalogue 
(Kube et al.\ \cite{kube+02}) records only one out of twenty confirmed IPs 
below the period gap. 

Finally, the non-uniform emission distribution in GZ Cnc marks it as an 
interesting system for further and more detailed studies using higher resolved 
spectroscopic data.

\begin{acknowledgements}
We thank Boris G\"ansicke and Ronald Mennickent for advice and discussions.
John Thorstensen kindly informed us on the result of Bill Fenton's and his
research, which is gratefully acknowledged. We thank the referee, Pablo 
Rodr\'{\i}guez-Gil, for helpful comments and for pointing out the behaviour of
high-velocity components in intermediate polars and SW Sex stars. 
This research has made use of the SIMBAD database, operated at CDS, Strasbourg,
France.
\end{acknowledgements}

\end{document}